\documentclass{blois}

\def\Journal#1#2#3#4{{#1} {\bf #2}, #3 (#4)}

\def\NIMA{{\em Nucl. Instrum. Methods} A}

\def\PLB{{\em Phys. Lett.}  B}
\def\PRL{\em Phys. Rev. Lett.}
\def\PRD{{\em Phys. Rev.} D}

\def\JHEP{{\em JHEP}}
\def\JINST{{\em JINST}}
\def\EPJC{{\em Eur. Phys. J.} C}

\def\be{\begin{equation}}
\def\ee{\end{equation}}
\def\bea{\begin{eqnarray}}
\def\eea{\end{eqnarray}}

\usepackage{amssymb}

\newcommand{\Photo}{}

\begin{document}
\vspace*{4cm}
\title{STUDY OF THE TOP QUARK PRODUCTION IN COMPLEMENTARY \\ PHASE SPACE REGIONS AND IMPACT ON PDFs IN CMS}

\author{ GEORGIOS KONSTANTINOS KRINTIRAS \\ on behalf of the CMS Collaboration }

\address{Universit\'e catholique de Louvain, Louvain-la-Neuve, Belgium\\
	 gkrintir@cern.ch}

\maketitle\abstracts{
The first measurement of the top quark pair production cross section ($\sigma_{\rm{t}\bar{\rm{t}}}$) in proton-proton collisions at $\sqrt{s} = 5.02$ TeV is reviewed. 
The data have been collected by the CMS experiment at the LHC and analyzed considering events with at least one charged lepton.
The extraction of $\sigma_{\rm{t}\bar{\rm{t}}}$
can be used to constrain the gluon distribution function (PDF) at large longitudinal parton momentum fraction and to establish experimentally
the relation between the top-quark mass as implemented in Monte-Carlo generators and the Lagrangian mass parameter.
The impact of the measurement on the determination of the gluon PDF is illustrated through a quantum chromodynamic analysis at next-to-next-to-leading order 
and the result is furthermore put in context with other top quark measurements in different phase space regions.
The measurement has paved the way for the first observation of top quark production in nuclear collisions and the subsequent 
study of modifications induced on the bound gluon PDF.}

\section{Introduction}
The top quark, the heaviest elementary particle in the standard model (SM),
is mainly produced in pairs (t$\bar{\rm{t}}$) via gluon-gluon fusion, at the LHC. 
The pair production cross section ($\sigma_{\rm{t}\bar{\rm{t}}}$) as a function of center-of-mass energy 
can be used to extract the top quark mass in a particular renormalization scheme and to constrain the gluon distribution function (PDF) 
at large fraction $x$ of the longitudinal proton momentum carried by a parton~\cite{rojo1,rojo2}, where the gluon PDF is poorly known.
Due to the large top quark mass ($m_{\rm{t}}$), 
the Yukawa coupling of the top quark with the Higgs boson is almost unity, 
hinting to a special role of the top-Higgs Yukawa interaction in the mechanism of electroweak symmetry breaking~\cite{buttazzo}.

In November 2015, the LHC delivered proton-proton (pp) collisions at $\sqrt{s}=5.02$ TeV~\cite{jowett} 
producing an integrated luminosity of $27.4\pm0.6\ \rm{pb^{-1}}$~~\cite{LUM16001}. 
Measurements of t$\bar{\rm{t}}$ production at various $\sqrt{s}$ probe different values of $x$ 
and thus provide complementary information on the gluon PDF.
Owing to the improved resolution on $x$, stronger constraints are placed by double-differential $\sigma_{\rm{t}\bar{\rm{t}}}$ measurements 
probing the details of the t$\bar{\rm{t}}$ production dynamics. 
A measurement of $m_{\rm{t}}$ can be also performed by simultaneously 
fitting $\sigma_{\rm{t}\bar{\rm{t}}} (m_{\rm{t}})$ and $m_{\rm{t}}^{\rm{MC}}$ i.e., by quantifying (``calibrating'') the difference relative to 
Monte Carlo (MC) manifestations of $m_{\rm{t}}$ that contain not only hard interactions but also contributions from  initial  and  final  state  radiation, 
hadronization,  as well as underlying-event interactions. Future studies of $\sigma_{\rm{t}\bar{\rm{t}}}$ 
in nuclear collisions at the same nucleon-nucleon center-of-mass energy would profit from the
availability of a reference measurement~\cite{enterria}, without the need to extrapolate from measurements at different $\sqrt{s}$.

\section{First measurement of $\sigma_{\rm{t}\bar{\rm{t}}}$  at $\sqrt{s}=5.02$ TeV}\label{subsec:prod}
\begin{figure}[htp]
\centering
\includegraphics[width=0.3\textwidth]{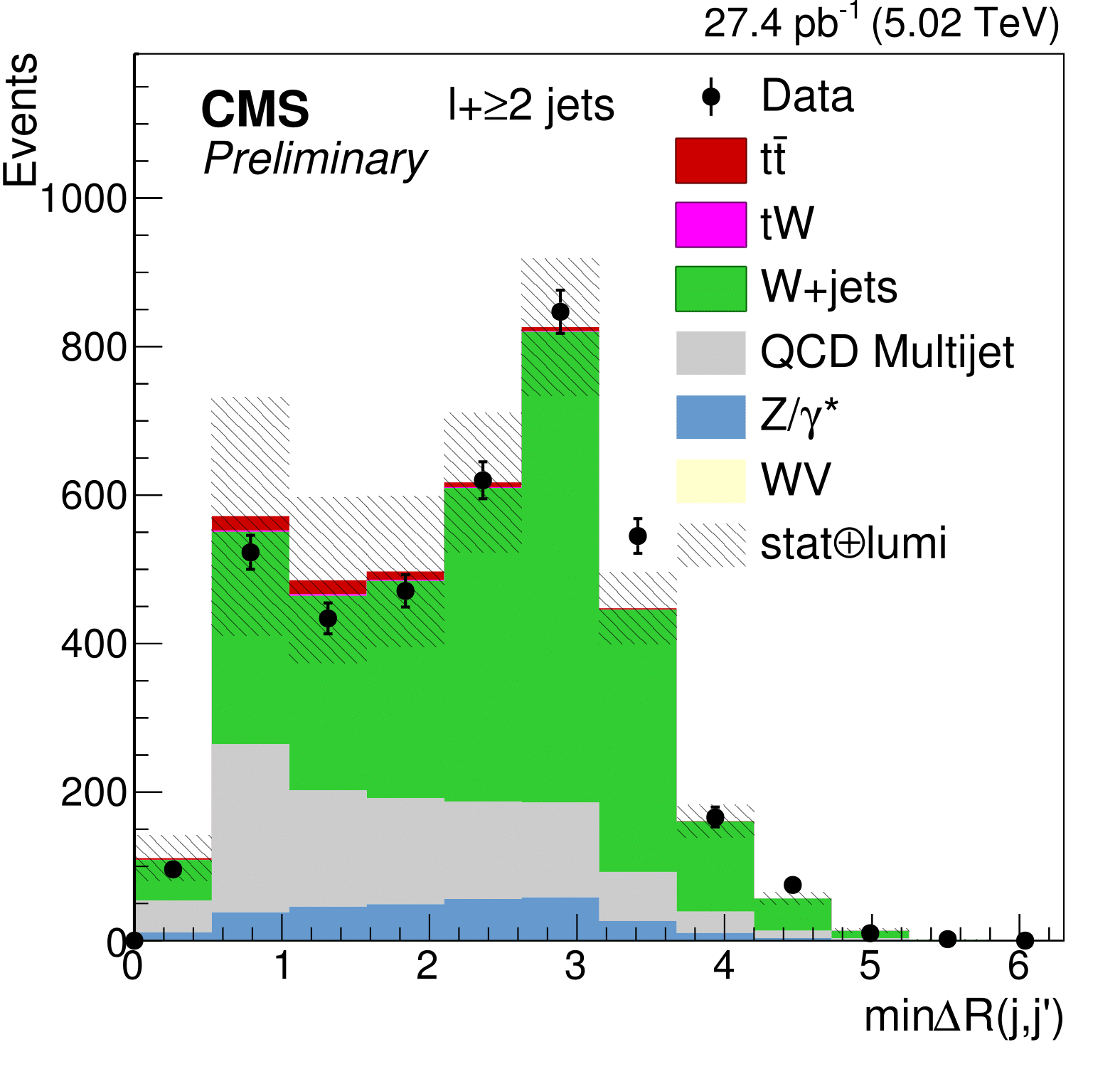}
\includegraphics[width=0.3\textwidth]{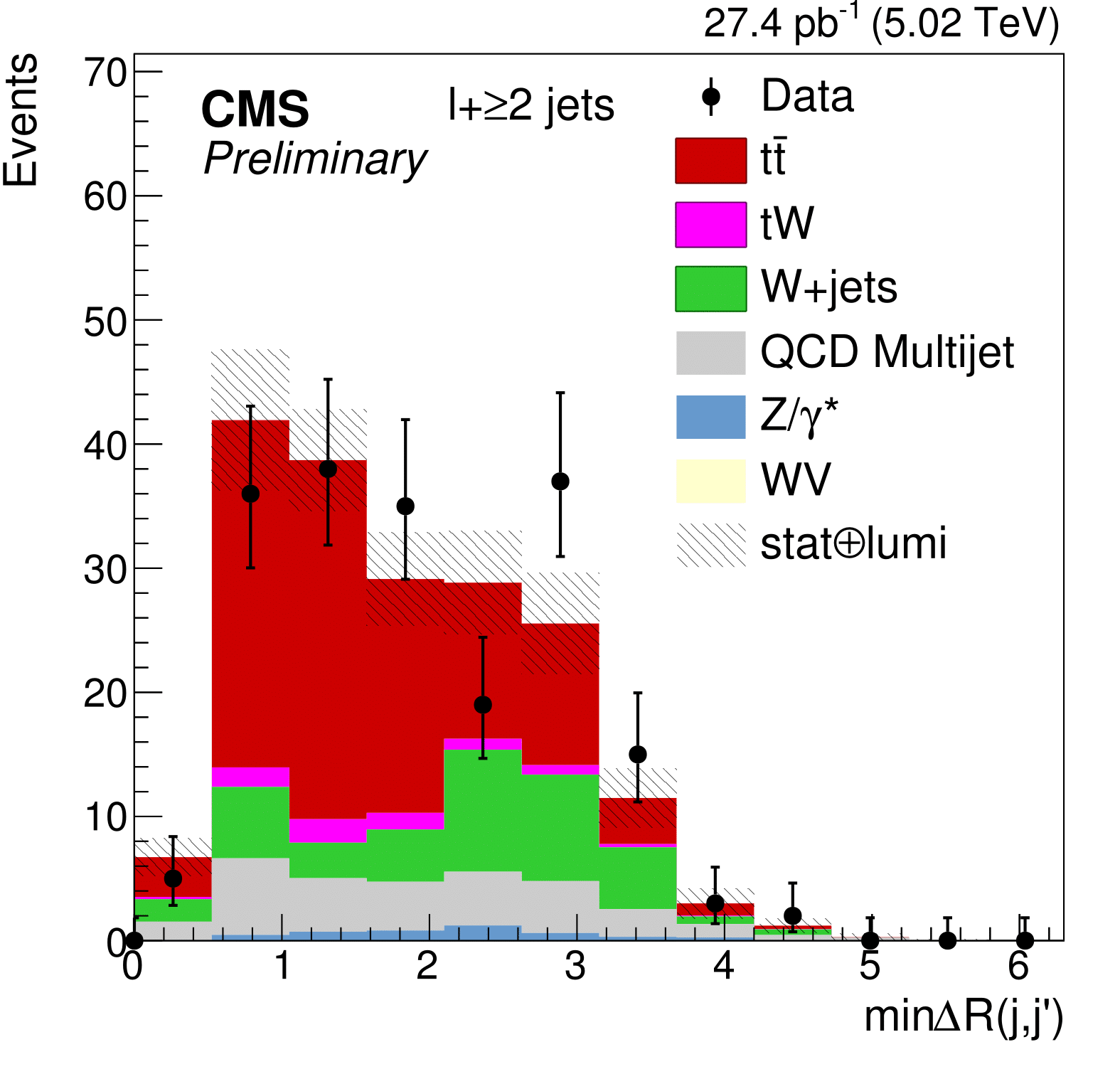}
\includegraphics[width=0.3\textwidth]{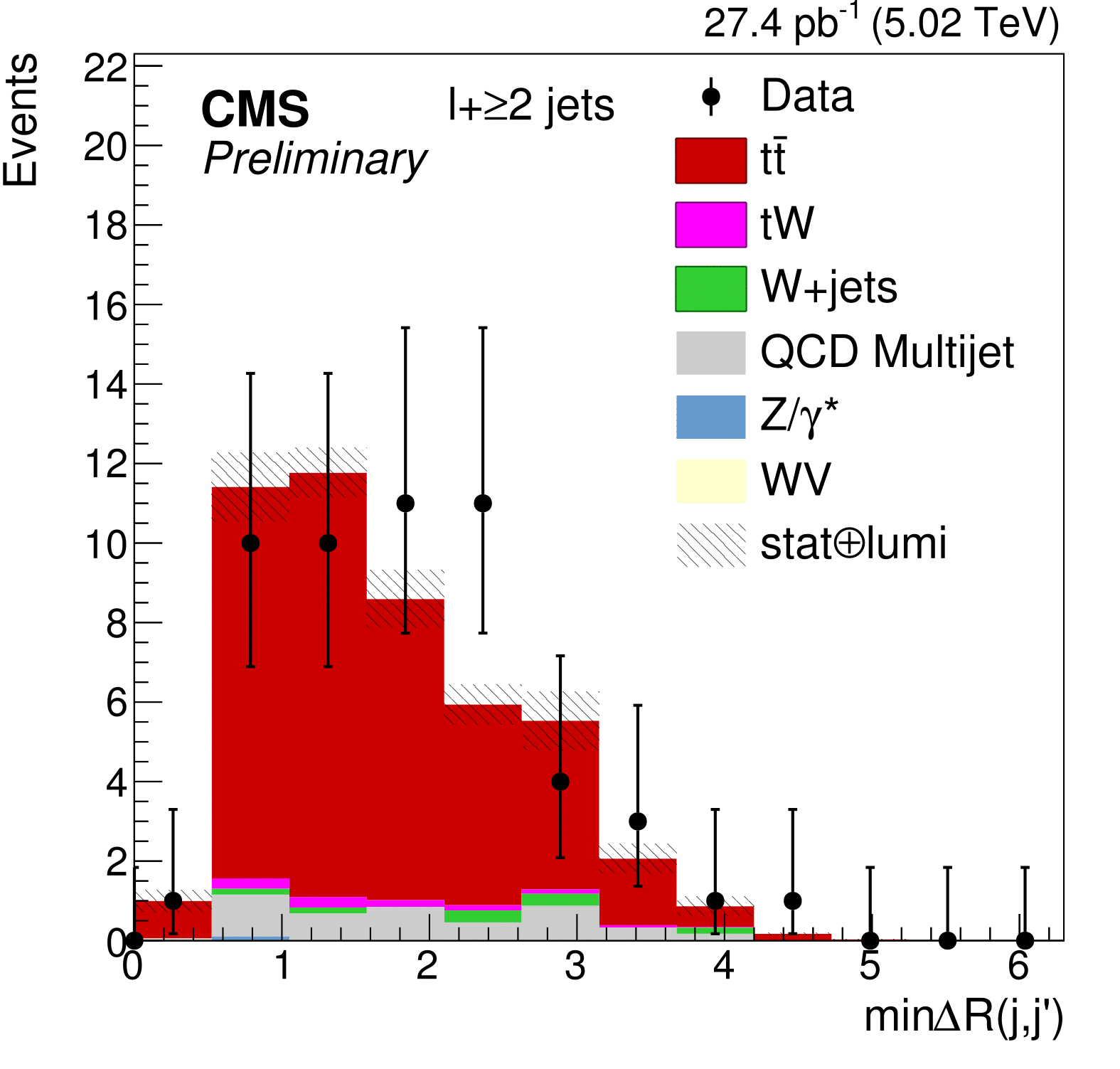}
\caption[]{Distribution of the angular distance $R=\sqrt{\eta^2+\phi^2}$ of the least separated jets, $\min\Delta R(j,j')$, for $\ell$+jets events
in 0b{}- (left), 1b{}- (center) and $\geq$2b{}- (right)  tagged jet categories.
The distributions observed in the data are compared to the sum of the expectations for the signal and backgrounds prior to the fit.
The shaded band represents the statistical and integrated luminosity uncertainties on the expected signal and background yields~\cite{TOP16023}.
}
\label{fig:ljets}
\end{figure}

Once produced, each top quark decays predominantly to a W boson and a bottom (b) quark. 
In the $\ell$+jets channel, one W boson decays leptonically and the other hadronically meaning the final state presents 
a typical signature of one isolated electron or muon, missing transverse momentum ($p_{\rm{T}}^{\rm{miss}}$), two jets coming from the hadronization of the b quarks (``b jets''), 
and two jets from the W boson hadronic decay (j,j$'$ jets). 
The correlation in phase space of these light jets carries a distinctive hallmark with respect to
the main backgrounds that are controlled by counting the number of b jets (``b tags'') in the selected events.
The signal extraction is then performed by a fit to the distribution of a kinematic variable~(Fig. \ref{fig:ljets}), sensitive to the resonant behavior of the light jets,  
for different categories of lepton flavor and jet multiplicity.

\begin{figure}[htbp!]
\centering
\includegraphics[width=0.4\textwidth]{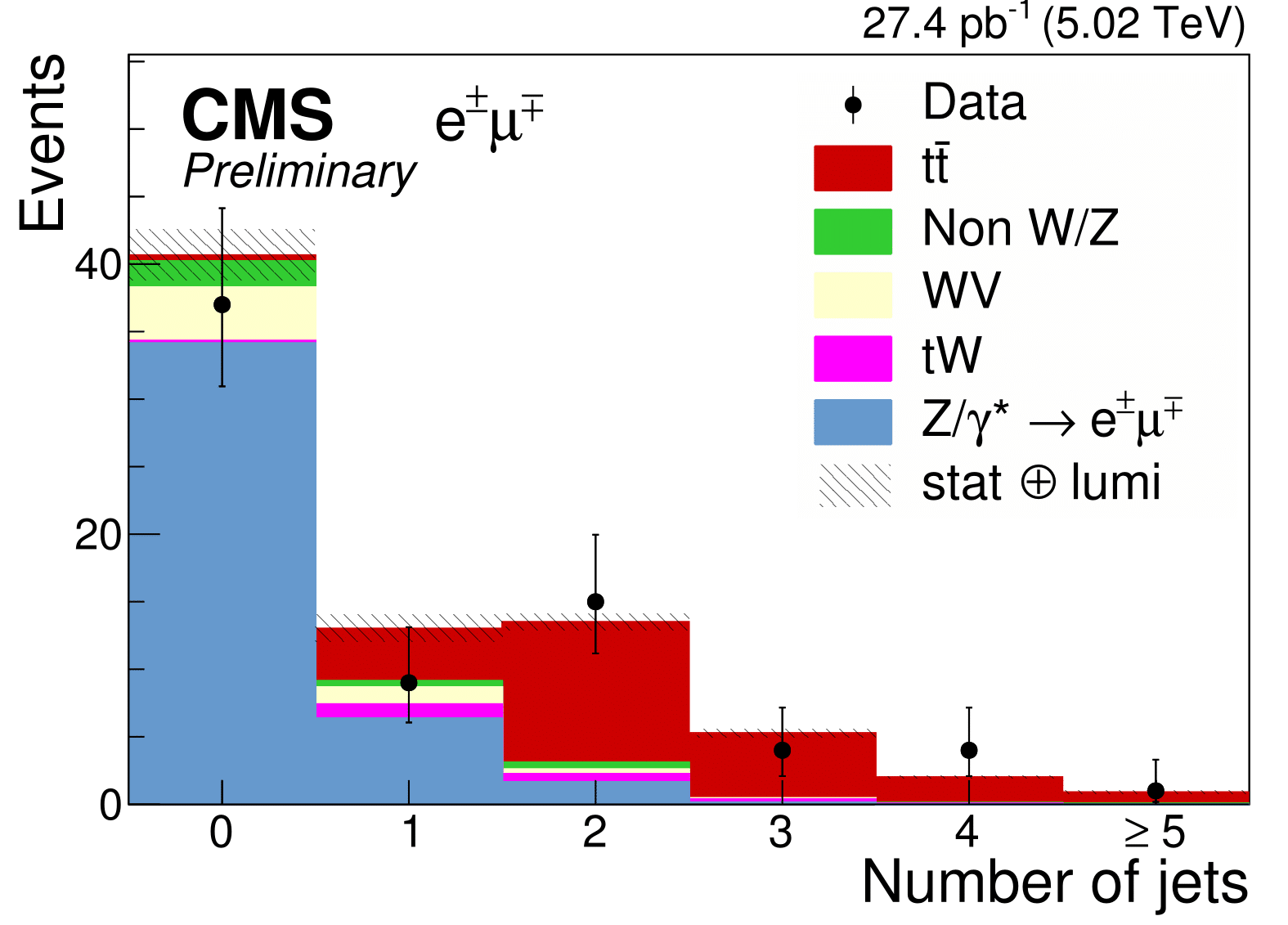}
\includegraphics[width=0.4\textwidth]{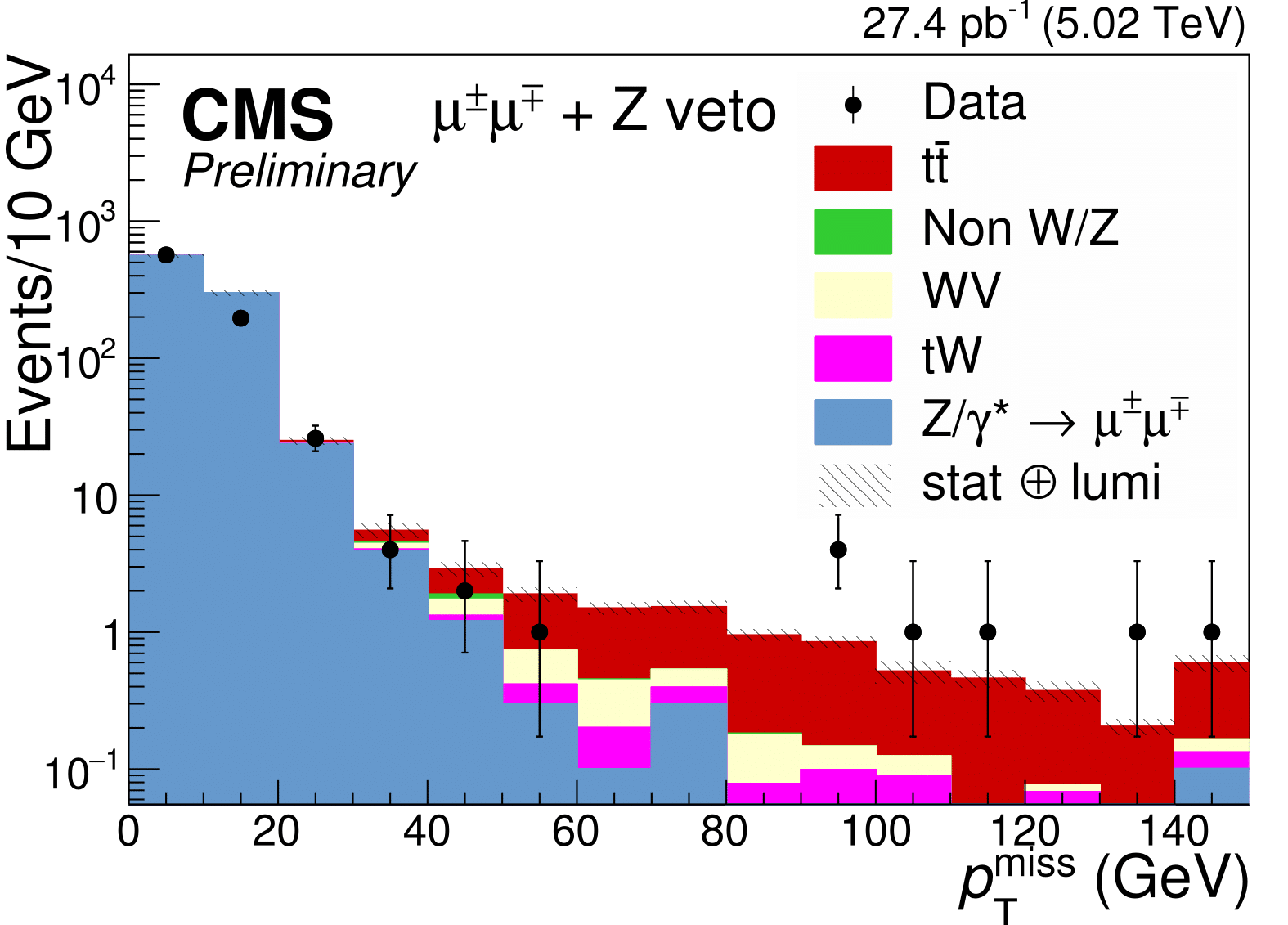}

\caption[]{Distributions of the (left) jet
multiplicity and of the (right) missing transverse momentum in events passing the selection criteria in the $\rm{e}^\pm \mu^\mp$ and 
$\mu^\pm \mu^\mp$ channels, respectively.
The shaded band represents the statistical and integrated luminosity uncertainties on the expected signal and background yields.
The last bin of the distributions contains the overflow events~\cite{TOP16023}. 
}
\label{fig:dilepton}
\end{figure}

On the other hand, in the dilepton channel, the final state contains two leptons
of opposite electric charge, $p_{\rm{T}}^{\rm{miss}}$, and at least two b jets. Although the $\ell$+jets
channel has a large branching ratio with a moderate amount of background, the dilepton
channel is characterized by a high purity, thus compensating for its smaller branching ratio even with a simpler yet robust event counting approach~(Fig. \ref{fig:dilepton}).

The individual results are then combined by using the BLUE method~\cite{lyons}, and
the robustness of the combination is tested by performing
an iterative variant of the combination method~\cite{lista} and varying some assumptions on the correlations of a series of systematics. 
The post-fit correlations between the nuisance parameters in the $\ell$+jets channel have been checked and found to introduce negligible impact.
The inclusive cross section is measured to be~\cite{TOP16023}:
\begin{eqnarray*}
 \sigma_{\rm{t}\bar{\rm{t}}} & = & 69.5 \pm 6.1~({\rm stat}) \pm 5.6~({\rm syst}) \pm 1.6~({\rm lumi})~{\rm pb}~=~69.5~\pm~8.4~({\rm total})~{\rm pb},
\end{eqnarray*}
with the weights of the individual measurements as 81.8\% for $\ell+$jets, 13.5\% for $\rm{e}\mu$ and 4.7\% for $\mu\mu$ channels. 
The achieved total relative uncertainty of $12\%$ represents a significant improvement relative to the first observation using the $\rm{e}\mu$ channel~\cite{TOP16015}.

\begin{figure*}[htbp]
\centering
{\includegraphics[width=0.5\textwidth]{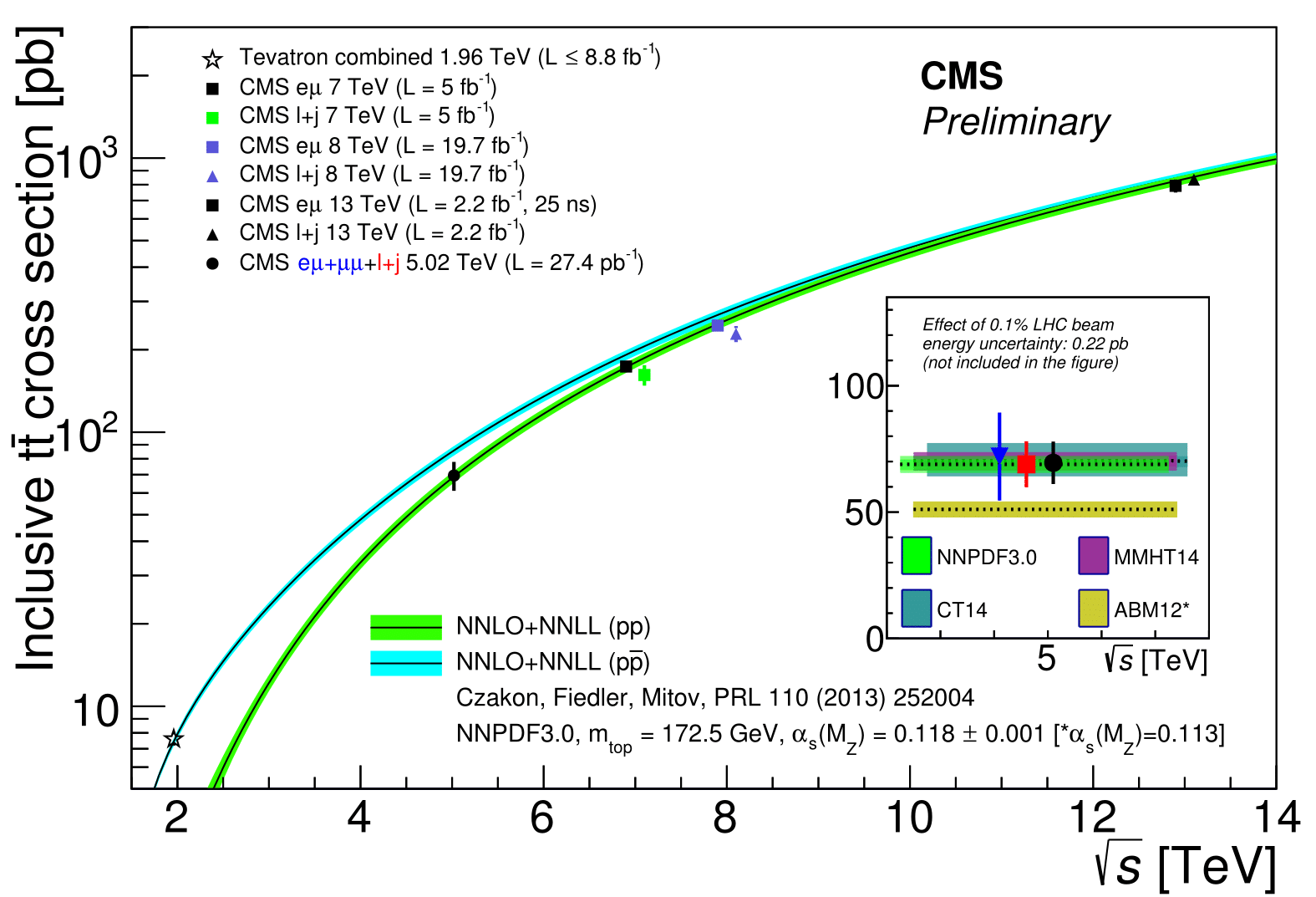}}
\caption[]{Top quark pair production cross section in
  pp and {\ensuremath{\rm p\bar{p}}} collisions as a
  function of the center-of-mass energy; the Tevatron combination at $\sqrt{s}= 1.96$ TeV~\cite{Tevatroncombination} is
  displayed, as well as CMS results at 5.02~\cite{TOP16023}, 7, 8~\cite{CMS7and8emu,CMS7and8ljets} and 13~\cite{CMS13dilep,CMS13ljets} TeV in the dilepton and $\ell$+jets channels.
  The measurements are compared to the NNLO+NNLL theory predictions~\cite{nnlonnll}.
}
\label{fig:sqrt}
\end{figure*}

Figure~\ref{fig:sqrt} presents a summary of results for $\sigma_{\rm{t}\bar{\rm{t}}}$ at various center-of-mass energies, including the
combination of the Tevatron measurements at
$\sqrt{s}=1.96$ TeV~\cite{Tevatroncombination} and the most precise CMS~\cite{CMS} measurements at
7 and 8 TeV~\cite{CMS7and8emu,CMS7and8ljets}, and 13 TeV~\cite{CMS13dilep,CMS13ljets}, compared to the theoretical predictions~\cite{nnlonnll}
at next-to-next-to-leading order in perturbative QCD, 
including soft-gluon resummation at next-to-next-to-leading-logarithmic order (NNLO+NNLL),
as a function of $\sqrt{s}$ for pp and proton-antiproton  ({\ensuremath{\rm p\bar{p}}}) collisions using the NNPDF3.0 NNLO PDF set~\cite{nnpdf30}.
The result at  $\sqrt{s}=5.02$ TeV is also compared to the predictions 
for the MMHT14~\cite{mmht14}, CT14~\cite{ct14}, and ABM12~\cite{abm12} PDF sets. With the exception of the ABM12 PDF set, we notice that the results for the rest of PDF sets are
close to each other, both in central values and theoretical uncertainties. The agreement between
them can be traced back to a similar default value of strong coupling ($\alpha_{\rm s}(M_{\rm Z})$) and a similar large $x$ gluon PDF. 
On the other hand, regarding the differences with respect to the ABM12 PDF set, it can be ascribed to
the smaller value of $\alpha_{\rm s}$ used by ABM12 and the softer large $x$ gluon PDF in the relevant region for top quark production.

\subsection{QCD analysis at NNLO}

\begin{figure}[htp]
\center                                                                                                                                           
   \includegraphics[width=0.3\textwidth]{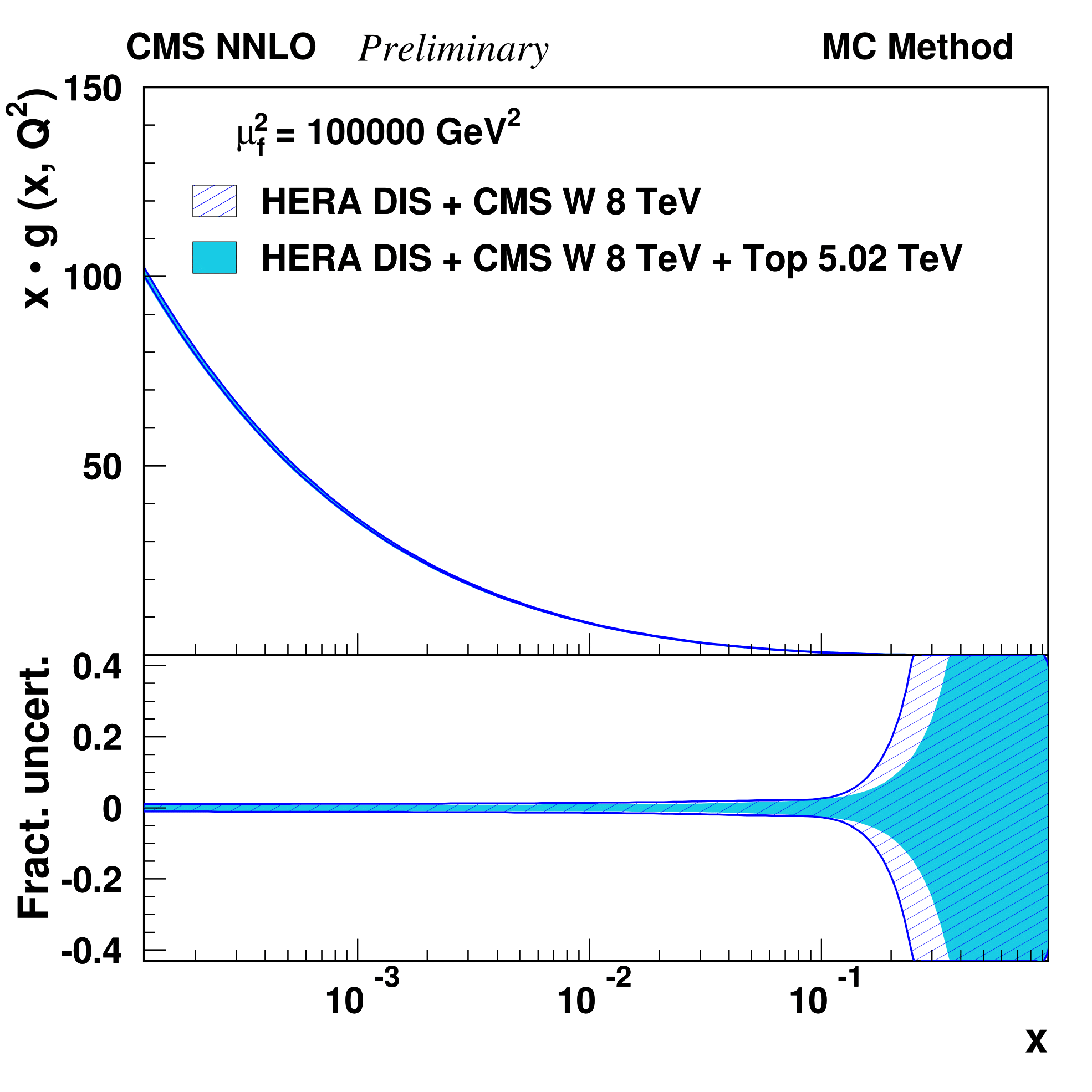}
   \includegraphics[width=0.3\textwidth]{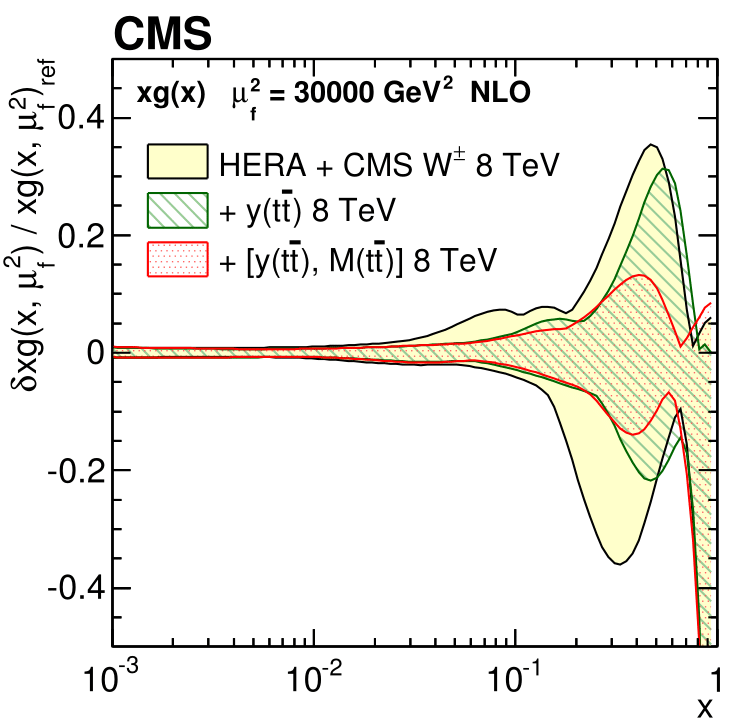}

\setlength{\unitlength}{1cm}
\caption[]{The gluon distribution and relative fractional uncertainties as functions of $x$ at the scale of (left)~\cite{TOP16023} 100000 and (right)~\cite{CMSdoublediff} 300000 GeV$^2$. 
The results of the fit including top-quark measurements, and without those are compared.}
\label{fig:PDFs}
\end{figure}

The impact of the $\sigma_{\rm{t}\bar{\rm{t}}}$ measurement at $\sqrt{s}=5.02$ TeV on the knowledge of the
proton PDFs is illustrated trough a QCD analysis at NNLO, together with the combined
measurements of neutral- and charged-current cross sections for deep inelastic electron- and
positron-proton scattering (DIS) at HERA~\cite{HERACombined}, and the CMS measurement of the muon charge asymmetry in W boson production~\cite{CMSWasymm}.
The precise HERA DIS data, obtained from the combination of the individual H1 and ZEUS results, are directly sensitive
to the valence and sea quark distributions and probe the gluon distribution through scaling violations, rendering these data the core of all PDF fits.
Parametrized PDFs are fitted to the data in a procedure that is referred to as the PDF fit and in that case version 2.0.0 of $\sc{\rm{xFITTER}}$, 
the open-source QCD-analysis framework for PDF determination~\cite{xFITTER}, is used.  
The experimental uncertainties in the measurement are propagated to the 14-parameter fit using the MC method.
Indeed, a moderate reduction of the uncertainty in the gluon PDF at high $x\gtrsim 0.1$ is observed, once
the measured values for $\sigma_{\rm{t}\bar{\rm{t}}}$ are included in the fit~(Fig. \ref{fig:PDFs}, left).  
All changes in the central values of the PDFs are well within the fit uncertainties, 
while the uncertainties in the valence quark distributions remain unaffected.

\section{Impact on PDFs and calibration of $m_{\rm{t}}^{\rm{MC}}$}

Precise measurements of $\sigma_{\rm{t}\bar{\rm{t}}}$ as a function of t$\bar{\rm{t}}$ kinematic 
observables have become possible, which allow both for the validation of the state-of-the-art QCD predictions and for a 
more precise determination of the gluon PDF at large values of $x$. 
The analysis~\cite{CMSdoublediff} uses the data recorded at $\sqrt{s}=8$ TeV by the CMS experiment in 2012, 
corresponding to an integrated luminosity of $19.7\pm0.5\ \rm{fb^{-1}}$~~\cite{LUM13001},
and it is restricted to the $\rm{e}^\pm \mu^\mp$ channel for a pure t$\bar{\rm{t}}$ event sample 
negligibly contaminated from Drell-Yan quark-antiquark annihilation into lepton-antilepton pairs through
Z boson or virtual photon exchange (``Z/$\gamma^*$''). 
Double-differential normalized $\sigma_{\rm{t}\bar{\rm{t}}}$ are used in a PDF fit at NLO, 
together with the DIS at HERA and the CMS W$^{\pm}$ boson charge asymmetry measurements. 
The uncertainties in the gluon distribution at $x > 0.01$ are significantly reduced once the t$\bar{\rm{t}}$ data
are included in the fit. The largest improvement originates from the $[y(\rm{t}\bar{\rm{t}}), M(\rm{t}\bar{\rm{t}})]$ $\sigma_{\rm{t}\bar{\rm{t}}}$ by
which the total gluon PDF uncertainty is reduced by more than a factor of two at $x \simeq 0.01$~(Fig. \ref{fig:PDFs}, right).

\begin{figure}[htp]
\center                                                                                                                                           
   \includegraphics[width=0.35\textwidth]{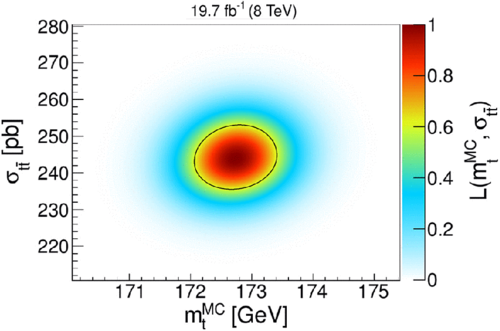}
   \includegraphics[width=0.35\textwidth]{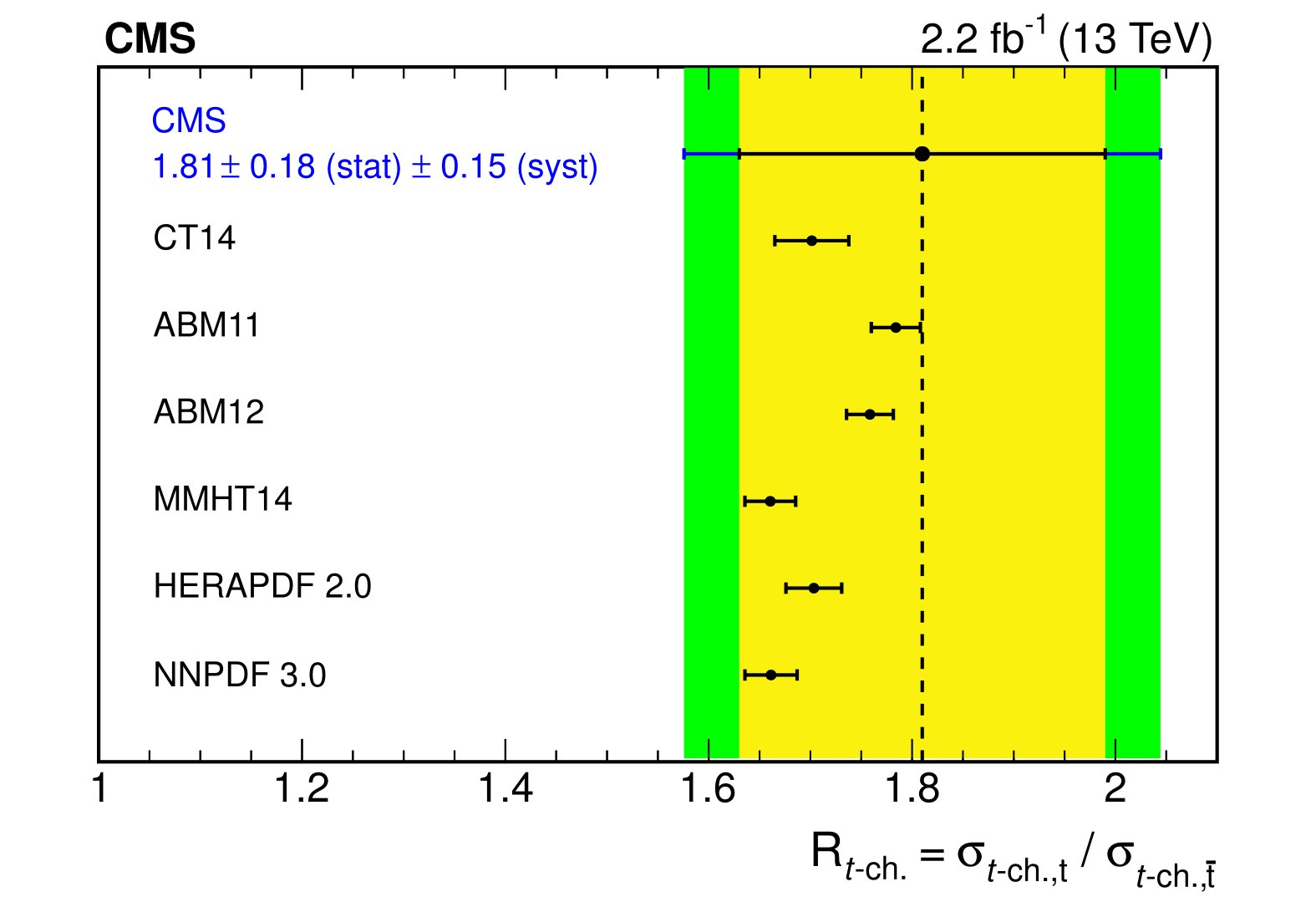}
\setlength{\unitlength}{1cm}
\caption[]{
Simultaneous likelihood fit for the (left)~\cite{mtopcalibration} measured $\sigma_{\rm{t}\bar{\rm{t}}} (m_{\rm{t}})$ and $m_{\rm{t}}^{\rm{MC}}$. 
The black contour corresponds to the 1 sigma uncertainty.
Comparison of the (right)~\cite{CMSsingletop} measured ratio of top quark and top antiquark production with the prediction from different PDF sets.
The error bars for the different PDF sets include the statistical uncertainty, 
the uncertainty due to the factorization and renormalization scales,
and the uncertainty in the top quark mass.
For the measurement, the inner and outer error bars correspond to the statistical and total uncertainties, respectively. 
}
\label{fig:ratioandfit}
\end{figure}

Currently, the translation from $m_{\rm{t}}^{\rm{MC}}$ to a theoretically well-defined mass definition can only be estimated to be of $\mathcal{O}(1)$ GeV,
resulting in the dominant systematic uncertainty in the $m_{\rm{t}}$ measurement which could be under- or overestimated.
For illustration, the measured $\sigma_{\rm{t}\bar{\rm{t}}}$,
using the data recorded at $\sqrt{s}=7$ and 8 TeV by the CMS experiment, is fitted simultaneously with $m_{\rm{t}}^{\rm{MC}}$.  
The measured values are mostly uncorrelated~(Fig. \ref{fig:ratioandfit}, left) and a calibration uncertainty of 2 GeV is achieved~\cite{mtopcalibration}.

\section{Outlook}

By including the top quark production data from LHC, a significant
reduction has been already observed in the uncertainties in the gluon distribution at large values of parton momentum
fraction $x$, in particular when using the double-differential $\sigma_{\rm{t}\bar{\rm{t}}}$ measurements. 
Dedicated analyses based on differential $\sigma_{\rm{t}\bar{\rm{t}}}$ seem to be a
promising approach to further decrease the uncertainty to measure well-defined $m_{\rm{t}}$ parameters.
Using measurements of the top quark production cross section in a global fit 
of the parton distribution functions can help to better determine both the gluon and valence quark distributions~(Fig. \ref{fig:ratioandfit}, right), while
the top quark is a novel probe of the nuclear gluon density at high virtualities in the unexplored high Bjorken-$x$ region~\cite{CMSttpPb}.


\section*{References}

\begin{thebibliography}{99}
\bibitem{rojo1}M. Czakon {\it et al}, \Journal{\JHEP}{07}{167}{2013}.

\bibitem{rojo2}M. Czakon {\it et al}, \Journal{\JHEP}{04}{044}{2017}.

\bibitem{buttazzo}D. Buttazzo {\it et al}, \Journal{\JHEP}{12}{089}{2013}.

\bibitem{jowett}J.M. Jowett {\it et al} in {\em The 2015 Heavy-Ion run of the LHC} (Proceedings of IPAC2016, Busan, Korea).

\bibitem{LUM16001}CMS Collaboration, CMS-PAS-LUM-16-001 (CMS Physics Analysis Summary, 2016, \url{https://cds.cern.ch/record/2235781}).

\bibitem{enterria}D. d'Enterria {\it et al}, \Journal{\PLB}{746}{64}{2015}.

\bibitem{lyons}L. Lyons {\it et al}, \Journal{\NIMA}{270}{110}{1988}.

\bibitem{lista}L. Lista, \Journal{\NIMA}{764}{82}{2014} ({\it Erratum}: \Journal{\NIMA}{773}{87}{2015}).

\bibitem{TOP16023}CMS Collaboration, CMS-PAS-TOP-16-023 (CMS Physics Analysis Summary, 2016, \url{https://cds.cern.ch/record/2258177}).

\bibitem{TOP16015}CMS Collaboration, CMS-PAS-TOP-16-015 (CMS Physics Analysis Summary, 2016, \url{https://cds.cern.ch/record/2161499}).

\bibitem{Tevatroncombination}CDF and D0 Collaborations, \Journal{\PRD}{89}{072001}{2014}.

\bibitem{CMS}CMS Collaboration, \Journal{\JINST}{3}{S08004}{2008}


\bibitem{CMS7and8emu}CMS Collaboration, \Journal{\JHEP}{08}{029}{2016}

\bibitem{CMS7and8ljets}CMS Collaboration, \Journal{\EPJC}{77}{15}{2017}

\bibitem{CMS13dilep}CMS Collaboration, \Journal{\EPJC}{77}{172}{2017}

\bibitem{CMS13ljets}CMS Collaboration, \Journal{\JHEP}{09}{051}{2017}

\bibitem{nnlonnll}M. Czakon {\it et al}, \Journal{\PRL}{110}{252004}{2013}.

\bibitem{nnpdf30}NNPDF Collaboration, \Journal{\JHEP}{04}{040}{2015}.

\bibitem{mmht14} L. A. Harland-Lang {\it et al}, \Journal{\EPJC}{75}{204}{2015}.

\bibitem{ct14} S. Dulat {\it et al}, \Journal{\PRD}{93}{033006}{2016}.

\bibitem{abm12} S. Alekhin {\it et al}, \Journal{\PRD}{89}{054028}{2014}.

\bibitem{HERACombined}ZEUS and H1 Collaborations, \Journal{\EPJC}{75}{580}{2015}.

\bibitem{CMSWasymm}CMS Collaboration, \Journal{\EPJC}{76}{469}{2016}.

\bibitem{xFITTER}$\sc{\rm{xFITTER}}$ Collaboration, http://www.xfitter.org.

\bibitem{CMSdoublediff}CMS Collaboration, \Journal{\EPJC}{77}{459}{2017}.

\bibitem{LUM13001}CMS Collaboration, CMS-PAS-LUM-13-001 (CMS Physics Analysis Summary, 2013, \url{https://cds.cern.ch/record/1598864}).

\bibitem{mtopcalibration}J. Kieseler {\it et al}, \Journal{\PRL}{116}{162001}{2016}.

\bibitem{CMSsingletop}CMS Collaboration, \Journal{\PLB}{772}{752}{2017}.

\bibitem{CMSttpPb}CMS Collaboration, \Journal{\PRL}{119}{242001}{2017}.




\end{thebibliography}
\end{document}